\documentclass[aps, prr, reprint, superscriptaddress, longbibliography]{revtex4-1}
\usepackage{eurosym}
\usepackage{graphicx}
\usepackage{epstopdf}
\usepackage{dcolumn}
\usepackage{bm}
\usepackage{here}
\usepackage[utf8]{inputenc}
\usepackage[T1]{fontenc}
\usepackage{mathptmx}

\begin{document}

\title{Elastic precursor effects during the Ba$_{1-x}$Sr$_{x}$TiO$_{3}$
ferroelastic phase transitions}
\author{Francesco Cordero}
\affiliation{Istituto di Struttura della Materia-CNR (ISM-CNR), Area della Ricerca di
Roma - Tor Vergata, Via del Fosso del Cavaliere 100, I-00133 Roma, Italy}
\email{[francesco.cordero@ism.cnr.it]}
\author{Francesco Trequattrini}
\affiliation{Dipartimento di Fisica, Universit\`{a} di Roma "La Sapienza", p.le A. Moro
2, I-00185 Roma, Italy}
\author{Paulo Sergio da Silva Jr.}
\affiliation{Department of Physics, Federal University of S\~{a}o Carlos,
13565-905 S\~{a}o Carlos (SP), Brazil}
\author{Michel Venet}
\affiliation{Department of Physics, Federal University of S\~{a}o Carlos,
13565-905 S\~{a}o Carlos (SP), Brazil}
\author{Oktay Aktas}
\affiliation{State Key Laboratory for Mechanical Behavior of Materials \& School of
Materials Science and Engineering, Xi'an Jiaotong University, Xi'an 710049,
China}
\author{Ekhard K. H. Salje}
\affiliation{Department of Earth Sciences, University of Cambridge, Downing Street,
Cambridge CB2 3EQ, UK}
\date{\today }

\begin{abstract}
Elastic softening in the paraelastic phases of Ba$_{1-x}$Sr$_{x}$TiO$_{3}$
is largest near the transition temperatures and decreases on heating
smoothly over extended temperature ranges. Softening extends to the highest
measured temperature(850~K) for Ba-rich compounds. The temperature
evolution of the excess compliance of the precursor softening follows a
power law $\delta S\propto |T-T_{\mathrm{C}}|^{-\kappa }$\ with a
characteristic exponent $\kappa $\ ranging between 1.5 in SrTiO$_{3}$\ and
0.2 in BaTiO$_{3}$. The latter value is below the estimated lower bounds of
displacive systems with three orthogonal soft phonon branches (0.5). An
alternative Vogel-Fulcher analysis shows that the softening is described by
extremely low Vogel-Fulcher energies $E_{a}$, which increase from SrTiO$_{3}$%
\ to BaTiO$_{3}$\ indicating a change from a displacive to a weakly
order/disorder character of the elastic precursor. Mixed crystals of Ba$_{x}$%
Sr$_{1-x}$TiO$_{3}$ possess intermediate behaviour. The amplitudes of the
precursor elastic softening increases continuously from SrTiO$_{3}$ to BaTiO$%
_{3}$. Using power law fittings reveals that the elastic softening is still $%
33\%$ of the unsoftened Young's modulus at temperatures as high as 750 K in
BaTiO$_{3}$ with $\kappa \simeq $ 0.2. This proves that the high temperature
elastic properties of these materials are drastically affected by elastic
precursor softening.
\end{abstract}

\maketitle




\section{Introduction}

Ferroelastic materials \cite{salje2012review} commonly display large elastic
anomalies during structural phase transitions \cite{carpenter1998}.
Structural collapses can lead to a total reduction of the effective moduli
in case of proper ferroelastics like in BiVO$_{4}$ and LaNbO$_{3}$ \cite%
{ishibashi1988, ishidate1989, errandonea1980}. In the case of improper
ferroelastics, like ferroelectric BaTiO$_{3}$\ and antiferrodistortive SrTiO$%
_{3}$, a direct coupling between the acoustic \cite{bussmannholder2009} soft
mode and the elastic moduli is symmetry forbidden while typical elastic
softening still reduces the moduli by some 20--50$\%$ \cite{kityk2000}. In
ferroelectrics, the intrinsic softening of the low-temperature phase with
respect to the paraelectric phase is due to the combined direct and
converse piezoelectric effects \cite{CCT16,Cor18b}. Additional softening in
the low temperature phases may be due to mobile twin boundaries just below
but close to the transition point, which vanishes if the twin walls are
strongly pinned \cite{salje2009cu, morozovska2013jap, he2022}. Thick domain
walls were shown to be less prone to such pinning effects \cite{lee2006,
lee2005, goncalves2010} and many examples of highly mobile wall movements
during the softening process have been reported \cite{ertas1996,
giamarchi1995, chrosch1999, wruck1994, harrison2002, klupfer1996}\emph{. }We
argue in this paper that significant precursor softening is commonly
observed in the paraelastic phase which can be \--- in some cases --
directly related to intrinsic disorder of the high temperature phase and
dynamic local nano-structures. Conceptually, this effect is best observed
when materials are disordered by extrinsic forces such as radioactive
bombardment. Consider a single crystal without any domain boundaries which
is then disordered by the radioactive decay of radiogenic impurities. Such
samples will massively reduce their elastic moduli due to the structural
heterogeneity. This situation is often encountered in so-called metamict
materials such as zircon \cite{salje2006zircon} and titanite \cite%
{salje2011rad} where the reduction in bulk and shear modulus is greater than
$50\%$, while their structure is still unchanged. Structural disorder of the
paraelastic phase and significant short-range order is, thus, expected if
the ferroelastic phase transition is of the order-disorder type and
structural variations occur in nominally cubic materials \cite{ZDW20,
aktas2021piezo, aktas2013pst, aktas2022review, zhao2022intrinsic}.

Displacive systems show similar effects although to a lesser extent \cite%
{cao1988}. Following the initial theoretical analysis of Pytte \cite%
{pytte1970, pytte1971} and Axe \& Shirane \cite{axe1970}, fluctuation
contributions to elastic softening have usually been considered in terms of
coupling between different vibrational modes \cite{hochli1972, rehwald1973,
cummins1979, luthi1981ultrasonic, yao1981, fossum1985}. In both cases local
fluctuations with high local correlations can be expected. Typical examples
for local nano-structures are tweed structures with interwoven, dynamical
strain fields. They were found by molecular dynamics simulations and
diffractions experiments \cite{SP91,MHN91,LCP08,NK09}. Similar effects were
postulated by Pelc \textit{et al.} \cite{PSA22} for high temperature
superconductors.

The underlying physical picture for displacive precursor softening is that,
associated with a soft mode at some specific point in reciprocal space,
there will be a set of branches which also soften to some extent. Along with
the soft mode itself, when the frequencies of modes along the soft branches
decrease, their amplitudes become larger and they can combine to produce
stress fluctuations and associated strain fluctuations. The summation of all
such combinations will yield a net softening of some specific acoustic modes
depending on the dimensionality of the elastic softening \cite{carpenter1998}%
. The total effect increases as the amplitudes of the modes increase,
reaching a maximum at the transition point. The detailed temperature
dependence is usually uncertain because it requires an exact knowledge of
mode mixing properties \cite{SP67}, including potential local modes which
are not captured by conventional spectroscopy. Nevertheless, it was
advocated that the resulting temperature dependence of the elastic softening
can be described conveniently by a power law \cite{carpenter1998,
wang2018,cao1988}:
\begin{equation}
\Delta C_{ik}=A_{ik}(T-T_{c})^{\kappa }.
\end{equation}%
$A_{ik}$ and $\kappa $ are properties of the material of interest. The
temperature $T_{c}$ is below the transition temperature $T_{\text{tr}}$.
Such power laws are inspired by simple soft-mode mechanisms, although $%
\kappa $ is sensitive to the details of mode-mode coupling, the degree of
anisotropy of dispersion curves about the reciprocal lattice vector of the
soft mode, and to the extent of softening along each branch. This effect
mimics the bilinear coupling between the order parameter and strain. In the
most common case, where direct interactions are symmetry forbidden, there
are still mechanisms which lead to softening in the displacive limit. For
example, optical phonons with opposite wavevectors $\mathbf{q}$ and $-%
\mathbf{q}$ can combine to produce a fluctuating strain field. The symmetry
allowed coupling is $e_{\mathrm{local}}\left\langle Q^{2}\right\rangle $,
where $e_{\mathrm{local}}$ is the local strain field and $\left\langle
Q^{2}\right\rangle $ is the average two-phonon amplitude.

Alternatively, when local disorder leads to thermally activated dynamics
with a broad distribution of activation energies and a threshold related to
the phase transition temperature, the elastic softening may follow a
Vogel-Fulcher statistics \cite{salje2013bto, carpenter2011pmn2} with\textit{%
\ }
\begin{equation}
\Delta C_{ik}=B_{ik}\exp \left( {\frac{E_{a}/k_{\text{B}}}{T-T_{\mathrm{VF}}}%
}\right) ,  \label{Eq:DCVF}
\end{equation}%
where $B_{ik}$ is a materials parameter, $E_{a}$ the activation energy and $%
T_{\mathrm{VF}}$ is the Vogel-Fulcher energy \cite{salje2014,
carpenter2011pmn2, thomson2014}. Even though Eq. (\ref{Eq:DCVF}) is not the
result of a formal theory, such behaviour is typical for glasses and for
local clusters in order-disorder phase transitions. In Ba$_{1-x}$Sr$_{x}$TiO$%
_{3}$ the two archetypal softening mechanisms, namely the displacive power
law and a mixed order-disorder Vogel-Fulcher mechanism can be expected to be
related to the antiferrodistortive displacive transition in SrTiO$_{3}$ at
105.6 K \cite{salje1998cubic} and the ferroelectric/ferroelastic transition
in BaTiO$_{3}$ at 400~K, which contains aspects of disorder from Ti
off-centering in the paraelectric phase \cite{zalar2003,zalar2005,Ber21}.

In materials investigated in this study the elastic precursor softening \cite%
{carpenter1998, kustov2020, migliori1993} is significant, including complex
intermediate cases \cite{Aub75,Aub76}. The power law softening and, to a
lesser extent, the Vogel-Fulcher softening describe the experimental
observations very well over a wide temperature range. The two limiting
cases, SrTiO$_{3}$ and BaTiO$_{3}$, behave very differently, and a smooth
variation of the model parameters $A$\emph{, }$\kappa $\emph{, }$B$\emph{, }$%
E_{a}$ is indeed observed in mixed crystals Ba$_{1-x}$Sr$_{x}$TiO$_{3}$ as
function of the chemical composition.

\section{Sample preparation}

Ceramic Ba$_{1-x}$Sr$_{x}$TiO$_{3}$ samples were prepared in two different
laboratories with different solid state sintering procedures. The BaTiO$_{3}$
sample is BT $\#1$ of Ref. \cite{cordero2021hopping}, prepared from
commercial high purity powder of BaTiO$_{3}$ ($99.9\%$, Sigma-Aldrich). The
powder was first heated at 800~K for 2~h to remove undesired organics and
ball milled for 24~h in order to reduce and homogenise the distribution of
particle sizes. It was then mixed with 3~wt.$\%$ polyvinyl butyral (PVB) as
a binder, uniaxially pressed at 150~MPa into a thick bar, isostatically
pressed at 250~MPa, and sintered at 1350~$^{\circ }$C for 2~h. The Ba$_{1-x}$%
Sr$_{x}$TiO$_{3}$ materials were synthesized starting from high-purity
powders of BaCO$_{3}$ ($99.8\%$, Alfa Aesar), SrCO$_{3}$ ($98\%$, Merck),
and TiO$_{2}$ ($99.8\%$, Merck) mixed for 24~h and then calcined at 1150~$%
^{\circ }$C for 5~h. The calcined BST powders were checked by XRD to be
perovskite tetragonal ($P4mm$) phase, milled for 24~h to obtain a
homogeneous particle size distribution around 1~$\mu $m, and mixed with 3~wt.%
$\%$ polyvinyl butyral (PVB) as a binder. Bar ingots were uniaxially pressed
at 190~MPa into metallic molds, isostatically pressed at 250~MPa and
conventionally sintered at 1350~$^{\circ }$C for 4~h. All samples were
checked by XRD and their densities are listed in Table \ref{table:density}.
\begin{table}[tbp]
\begin{tabular}{ccc}
x(Sr) & Density (g/cm\textsuperscript{3}) & Relative density ($\%$) \\
\hline\hline
0 & 5.64 & 93.7 \\
0.03 & 5.85 & 97.5 \\
0.1 & 5.83 & 97.9 \\
0.3 & 5.67 & 97.8 \\
0.5 & 5.35 & 95.9 \\
1 & 5.02 & 97.9 \\ \hline
\end{tabular}
\centering
\caption{Density of ceramic samples Ba$_{1-x}$Sr$_{x}$TiO$_{3}$, assuming a
linear theoretical density that changes linearly between 5.13~g/cm$^{3}$ of
SrTiO$_{3}$ and 6.02~g/cm$^{3}$ of BaTiO$_{3}$.}
\label{table:density}
\end{table}
The resulting ceramics were cut into thin bars with lengths of 35--42~mm,
widths of 4--6~mm and thicknesses of 0.4--0.7~mm, mechanically polished and
annealed at 750~$^{\circ }$C for 2~h to release stresses.

The SrTiO$_{3}$ sample was prepared by solid-state reaction of SrCO$_{3}$
(Aldrich, $99.9\%$) and TiO$_{2}$ (Aldrich, $99.9\%$) for 6~h at 1100~$%
^{\circ }$C. The resulting powder was milled, sieved and pressed in a bar
and sintered in air at 1450~$^{\circ }$C for 24~h. Thin bars 43~mm long and
0.5~mm thick were cut and polished for the anelastic measurements. The Ba$%
_{0.5}$Sr$_{0.5}$TiO$_{3}$ bar, 37~mm long and 0.44~mm thick, was prepared
in a similar manner. The sample of SrTiO$_{3}$\ had $T_{\text{tr}}\simeq 111$%
~K, which is 5.3 K above the value generally found. A possible explanation
is a larger than usual content of Ca impurities in the starting SrCO$_{3}$\
powder. To avoid artifacts due to impurities, we also used a single crystal
of SrTiO$_{3}$\ with dimensions of 26.15 3.4 0.5 mm$^{3}$, cut from a wafer
of M.T.I. Corporation with the edges parallel to the $\left\langle
100\right\rangle $\ directions.

\section{Experimental methods}

The dynamic Young's modulus $E$ was measured by electrostatically exciting
the free flexural modes of the bars suspended in vacuum on thin thermocouple
wires, as described in Ref. \cite{cordero2008insert}. The real part is
deduced from the resonant frequency of the fundamental flexural mode \cite%
{nowick1972}
\begin{equation}
f=1.028\frac{t}{l^{2}}\sqrt{\frac{E}{\rho }}.  \label{Eq:fres}
\end{equation}

For the $\left\langle 100\right\rangle $ oriented crystal of SrTiO$_{3}$ we
measured $E=s_{11}^{-1}$. We do not consider the absolute values of the
Young's moduli, but their normalized values, $E(T)/E_{0}=\left(
f(T)/f_{0}\right) ^{2}$, where the temperature dependence of the density in
Eq. (\ref{Eq:fres}) is ignored in comparison with the elastic moduli. Its
linear dependence does not influence the fits, since it is absorbed in the
fit of the background elastic modulus.

\section{Results}

\begin{figure}[t]
\begin{center}
\includegraphics[width=8.5cm]{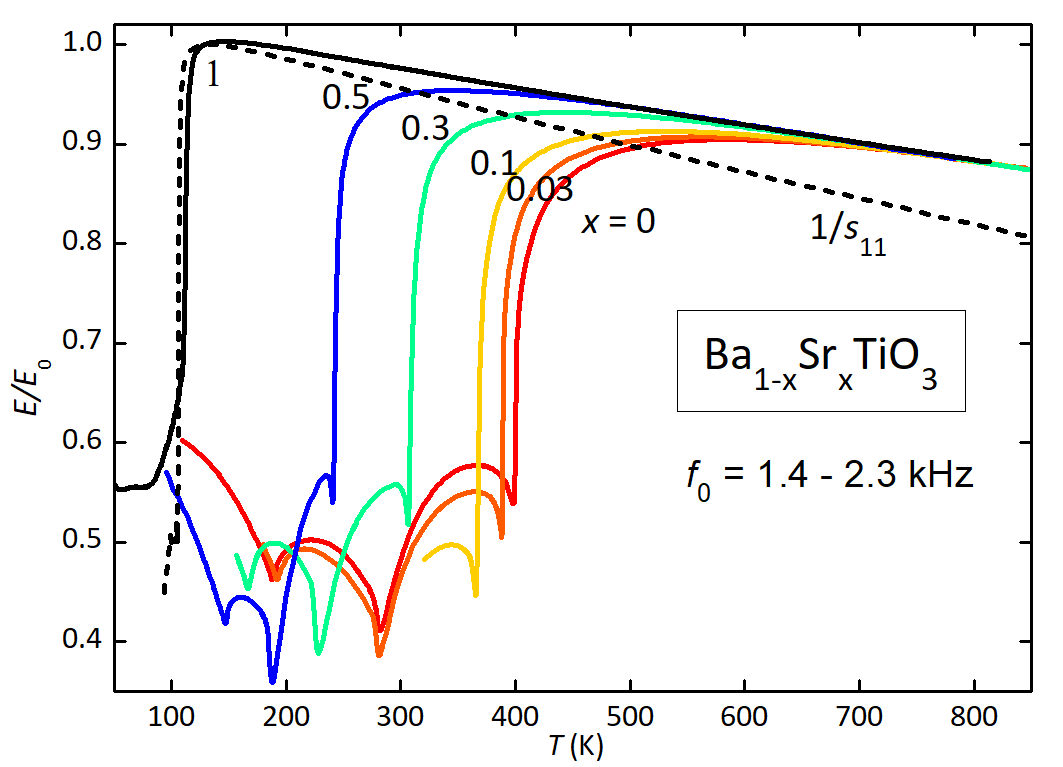}
\end{center}
\caption{Young's moduli of Ba$_{1-x}$Sr$_{x}$TiO$_{3}$ ceramics normalized
to overlap at $T\gg $ $T_{\mathrm{C}}$. The dashed line is $1/s_{11}$ of
the SrTiO$_{3}$ crystal.}
\label{fig:norm}
\end{figure}

Fig.~\ref{fig:norm} presents the Young's moduli of Ba$_{1-x}$Sr$_{x}$TiO$%
_{3} $ normalized to overlap at high temperature. The purpose of this figure
is to show that the anharmonic linear softening (\textit{i.e.} the slope $%
dE/dT$) is similar for all compositions. The slope of the single
crystal is different because the Young's modulus of a ceramic cubic material
is%
\begin{equation}
E^{-1}=S_{11}-\frac{2}{5}\left( S_{11}-S_{12}-\frac{1}{2}S_{44}\right) ,
\label{Eq:E}
\end{equation}%
while that of a $\left\langle 100\right\rangle$
oriented crystal is $E^{-1}=S_{11}$ \cite{nowick1972}, and the
various $S_{ij}$ may have different slopes.

\begin{figure}[t]
\begin{center}
\includegraphics[width=8.5cm]{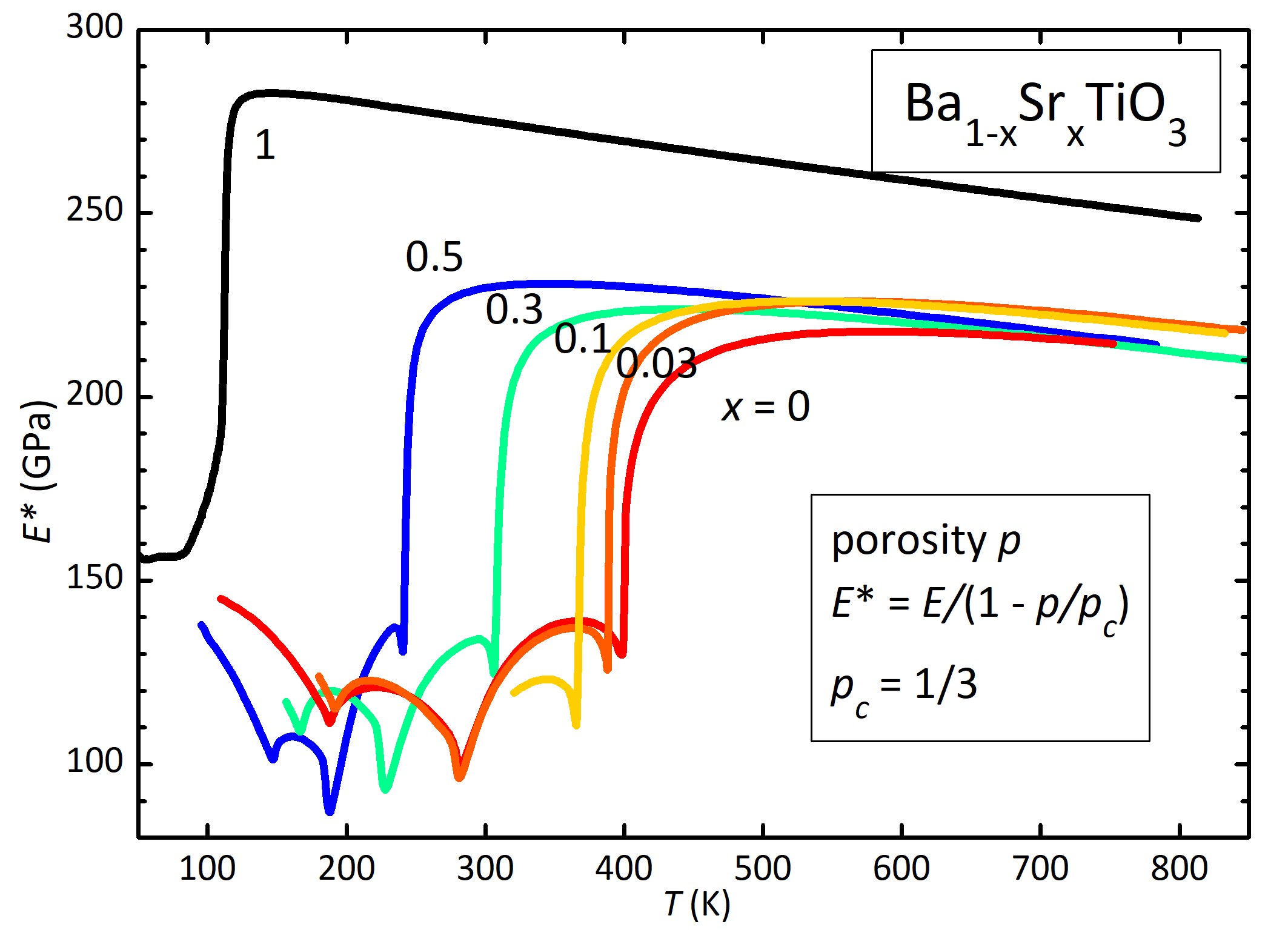}
\end{center}
\caption{Young's moduli of Ba$_{1-x}$Sr$_{x}$TiO$_{3}$ corrected for
porosity of the ceramic samples. }
\label{fig-absolute}
\end{figure}

The measurement of BaTiO$_{3}$ was already reported in Ref. \cite%
{cordero2021hopping} (Fig.~1, sample BT $\#1$). In addition to the step at
the transition from the cubic paraelectric to the tetragonal ferroelectric
phases, there are two additional minima due to the ferroelectric transitions
to the orthorhombic and rhombohedral phases. In SrTiO$_{3}$ the structural
transition is antiferrodistortive, with tilting of the TiO$_{6}$ octahedra
around the $c$ axis. A changeover of the character of the structural
transitions in Ba$_{1-x}$Sr$_{x}$TiO$_{3}$ was projected at $x\approx $
0.8--0.9 \cite{lemanov1996}. The normalization of $E$ in Fig. 1 only shows
that the temperature slopes in the high temperature limits of $E(T)$
coincide even though the absolute values of the moduli are different. These
absolute moduli are difficult to compare with each other due to
irregularities in the sample shapes and differences in porosity, which are
not included in Eq. (\ref{Eq:fres}) but reduce $E$ in manners that depend on
the amount and type of porosity \cite{salje2010minmag, dunn1995, cordero2018}%
. We corrected for porosity effects using the linear decrease in the Young's
modulus \cite{salje2010minmag},
\begin{equation}
E=E^{\ast }(1-p/p_{c}),  \label{Eq:porosity}
\end{equation}%
where $E$\ is the measured modulus, $E^{\ast }$\ the modulus of the dense
material and $p_{c}$ is a critical porosity. We chose $p_{c}=1/3$ because it
yields 218~GPa for the maximum value reached by BaTiO$_{3}$ in the PE phase,
in good agreement with 214~GPa for the single crystal \cite{cordero2018,
berlincourt1958}, and 275 GPa for the room temperature value of SrTiO$_{3}$,
in agreement with 285~GPa of Ref. \cite{carpenter2007b}. The result is shown
in Fig.~\ref{fig-absolute}.

The modulus of the paraelastic phase of Ba$_{1-x}$Sr$_{x}$TiO$_{3}$
increases with $x$, though the $x=$ 0.3 and 0.5 compositions are well below
a linear trend. This may be an artifact because the relative density
measured by the Archimedes method underestimates the open porosity, and
strong deviations from a simple formula like Eq. (\ref{Eq:porosity}) may
occur, as shown for BaTiO$_{3}$ \cite{cordero2018}.

\begin{figure}[t]
\begin{center}
\includegraphics[width=8.5cm]{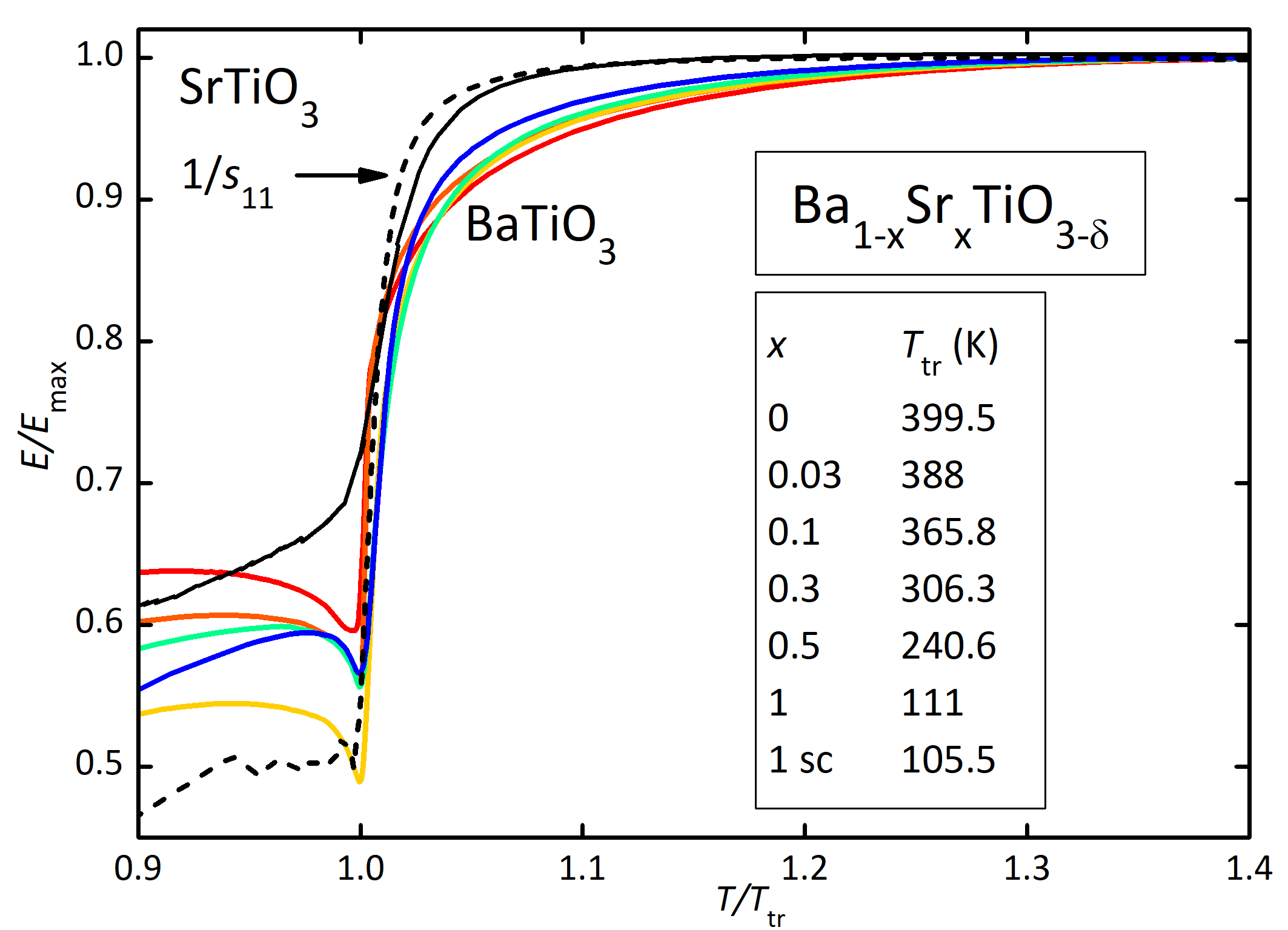}
\end{center}
\caption{Young's moduli of Ba$_{1-x}$Sr$_{x}$TiO$_{3}$ normalized to their
maximum on a $T/T_{tr}$ scale with $T_{tr}$ listed in the insert.}
\label{fig-NEvsNT}
\end{figure}

We show in Fig. \ref{fig-NEvsNT} the data normalized both in magnitude with
respect to the maximum value of the modulus, and temperature with respect to
$T_{\text{tr}}$ to emphasise the continuous change of the precursor
softening above $T_{\text{tr}}$ with composition, from a gradual decrease of
BaTiO$_{3}$ to a sharp decay in SrTiO$_{3}$.

For numerical fits, the compliances $S_{ij}$\ are split into the
background compliance $S_{ij}^{bg}$\ and the precursor softening $\Delta S$\
due to the phase transition. The temperature dependence of $S^{bg}$\ is
approximated to be linear with saturation below a temperature $\Theta $,
that can be described in terms of the quantum saturation expression \cite%
{SWT91}
\begin{equation}
S_{bg}=S_{0}+S_{1}\coth \left( \Theta /T\right) ~.  \label{Eq:bg}
\end{equation}

The precursor softening is
\begin{equation}
\Delta S_{\mathrm{power}}=A\left( T/T_{\mathrm{C}}-1\right) ^{-\kappa }
\label{Eq:pow}
\end{equation}%
for the power law and

\begin{equation}
\Delta S_{\mathrm{VF}}=B\left( 1-\exp \left[ -E_{a}/\left( T-T_{\mathrm{VF}%
}\right) \right] \right)  \label{Eq:VF}
\end{equation}%
for the Vogel-Fulcher approach. If all the compliances $S_{ij}$\ have the
same temperature dependences of background and precursor components, though
with different amplitudes, the fitting expression%
\begin{equation}
E/E_{0}=1/S=1/\left( S_{0}+S_{1}\coth \left( \Theta /T\right) +\Delta
S\right)  \label{Eq:fit}
\end{equation}%
is valid for both ceramics and single crystal.

\begin{figure}[t]
\begin{center}
\includegraphics[width=8.5cm]{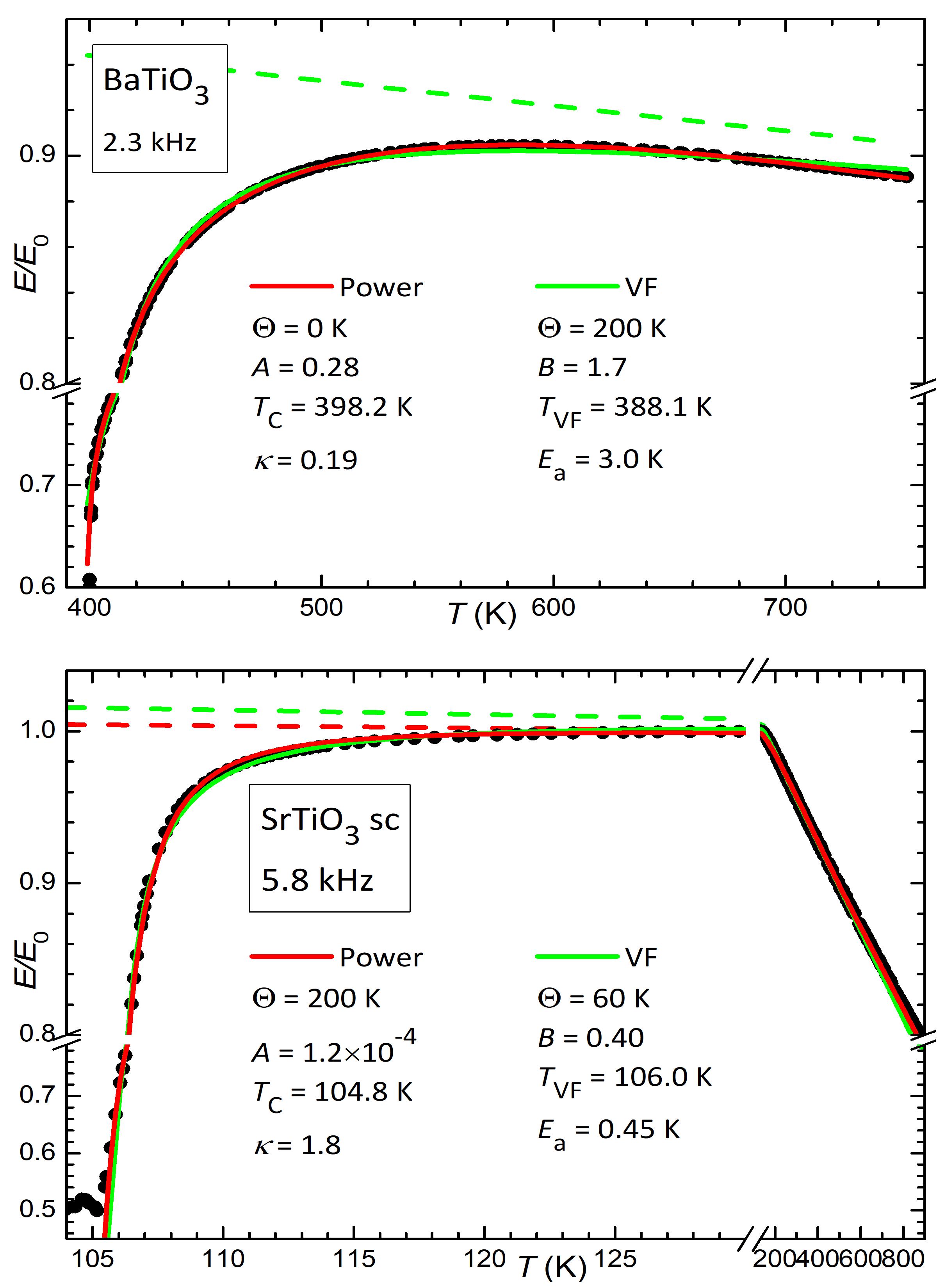}
\end{center}
\caption{Fits of the normalized Young's moduli of BaTiO$_{3}$ and SrTiO$_{3}$
with Eqs. (\protect\ref{Eq:fit}-\protect\ref{Eq:VF}) for the power law and
Vogel-Fulcher analyses and the parameters indicated in the legends. The
dashed lines represent $S_{bg}$, obtained setting $A$ or $B=0$. For the case
of the power law in BaTiO$_{3}$, $S_{bg}$ is out of scale: from 1.18 at
760~K to 1.38 at 400~K.}
\label{fig-fits}
\end{figure}

Figure \ref{fig-fits} shows fits of the normalized moduli of ceramic BaTiO$%
_{3}$ and single crystal SrTiO$_{3}$ with Eq. (\ref{Eq:fit}) and both Eqs. (%
\ref{Eq:pow}) and (\ref{Eq:VF}) and the parameters indicated in the legends.
The dashed lines represent the background $S_{bg}$, which is obtained
setting $A$ or $B=0$. For the case of the power law in BaTiO$_{3}$, $S_{bg}$
is out of scale: from 1.18 at 760~K to 1.38 at 400~K. Similar fits are
obtained for the intermediate compositions.

\begin{figure}[t]
\begin{center}
\includegraphics[width=8.5cm]{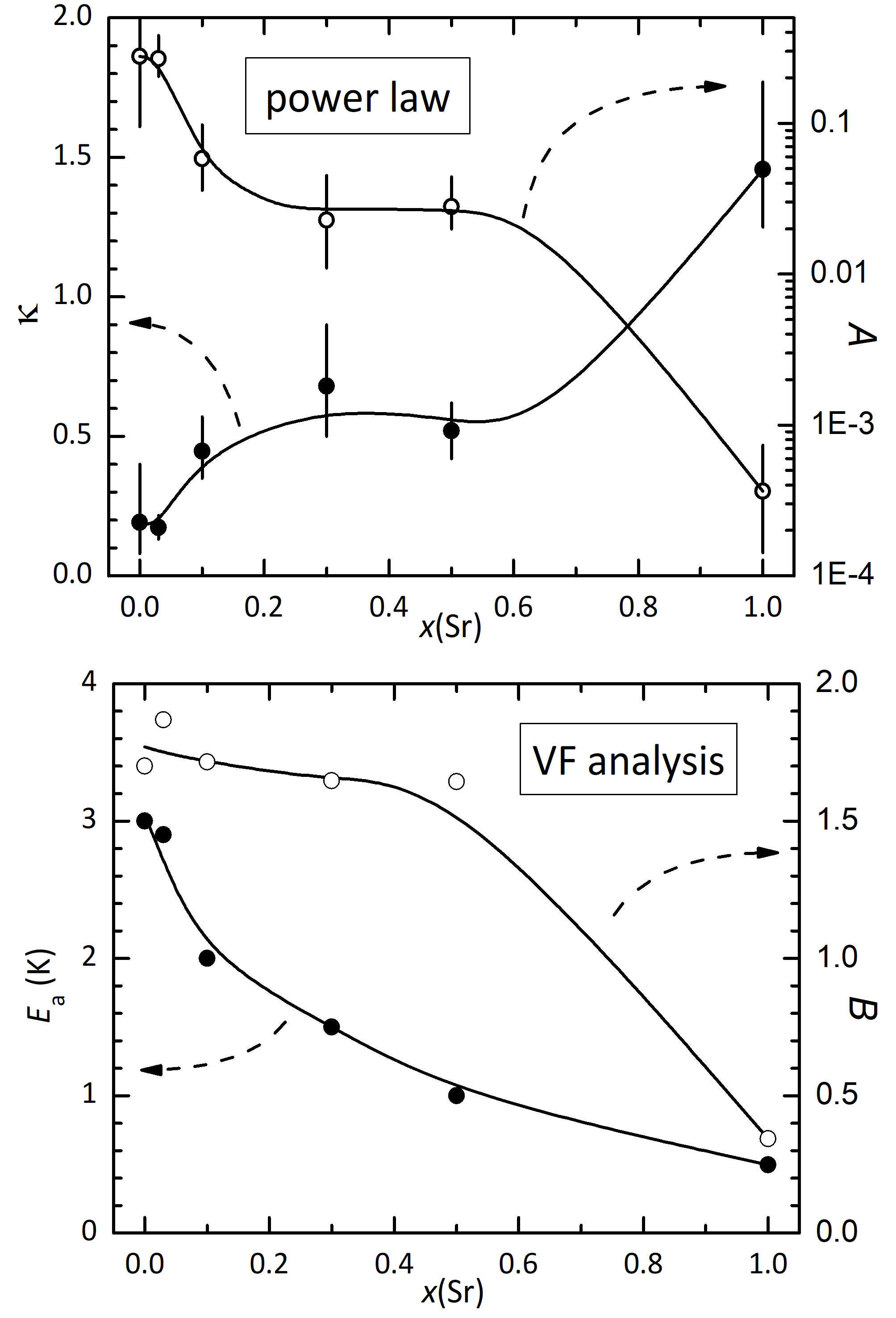}
\end{center}
\caption{Variation of power law and Vogel-Fulcher fit parameters with
composition $x$. The continuous lines are guides for the eye.}
\label{fig-param}
\end{figure}

Figure \ref{fig-param} shows the dependence of the parameters $\kappa ,A$
and $E_{a},B$ on composition $x$. The error bars indicate the parameters
regions where $\chi ^{2}/\chi _{\min }^{2}\leq 2$. The error bars for the VF
parameters are quite large, because the VF expression tends to $B\times
E_{a}/\left( T-T_{\mathrm{VF}}\right) $ for small $E_{a}/\left( T-T_{\mathrm{%
VF}}\right) $, and therefore the fit does not change reducing $E_{a}$ below
a certain value. This situation occurs at all compositions, where a decrease
but no definite minimum of $\chi ^{2}$ is obtained decreasing $E_{a}$ and
simultaneously increasing $B$, except for SrTiO$_{3}$, where a clear minimum
is found for $E_{a}=$ 0.50$_{-0.09}^{+0.05}$~K.

The power law dependence is well demonstrated with exponent $\kappa \simeq
0.2$ for BaTiO$_{3}$ increasing continuously to 1.5 for SrTiO$_{3}$. The
activation energy $E_{a}$ decreases from $\sim 3$~K to 0.50~K.
Simultaneously, the amplitudes $A$ and $B$ decrease continuously. For both
parameters a plateau for intermediate compositions near $x=$ $0.4$ appears
to occur.

\begin{figure}[t]
\begin{center}
\includegraphics[width=8.5cm]{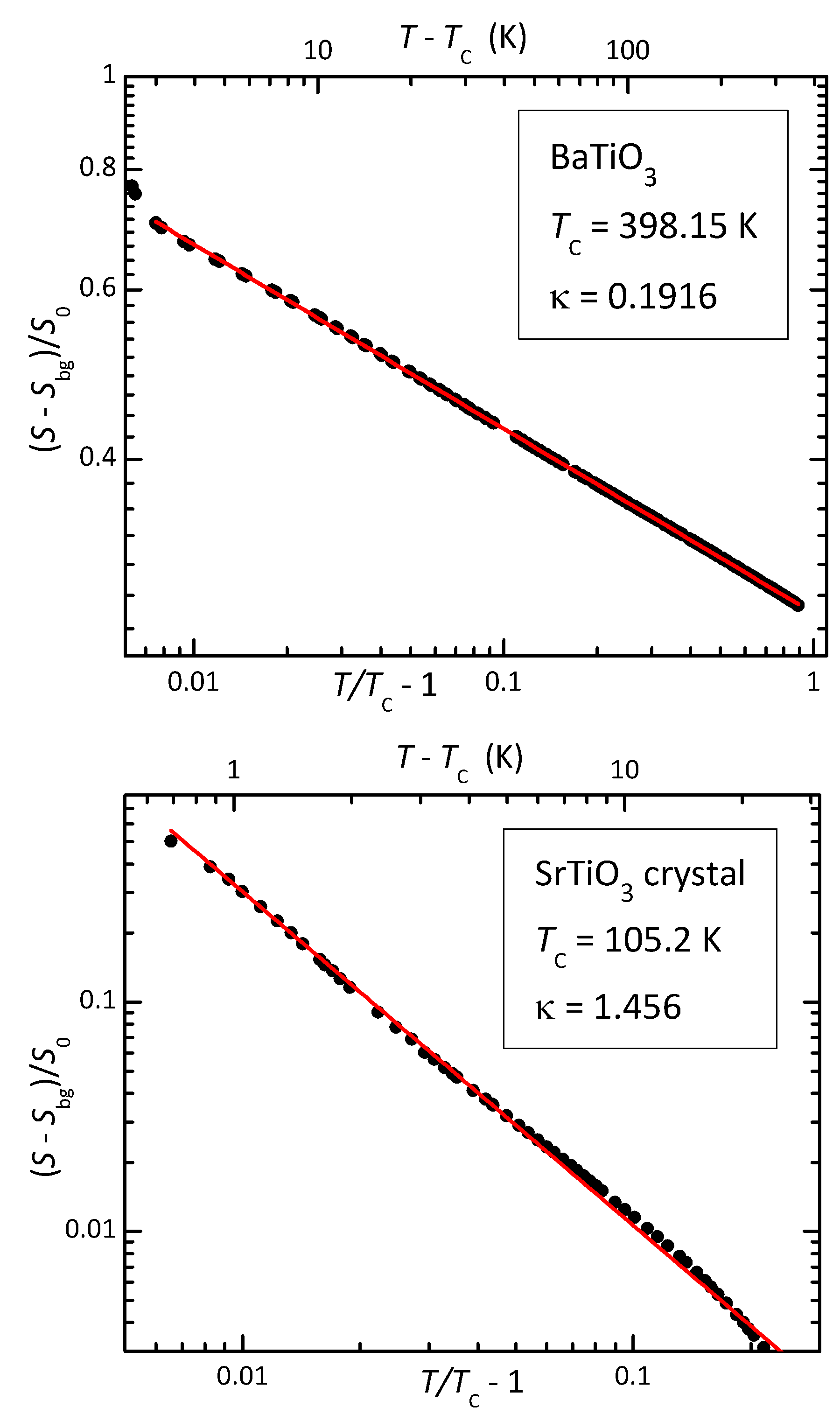}
\end{center}
\caption{Double logarithmic plots of the precursor softening vs reduced
temperature of BaTiO$_{3}$ and SrTiO$_{3}$, $\Delta S_{\mathrm{power}}=$ $%
\left( E/E_{0}\right) ^{-1}-S_{bg}$, where $S_{bg}$ is obtained from the
fits setting $A=0$. The real temperatures are shown in the upper scales.}
\label{fig-loglog}
\end{figure}

To better show the close adherence of the precursor softening of BaTiO$_{3}$
and SrTiO$_{3}$ to a power law, Fig. \ref{fig-loglog} presents double
logarithmic plots versus reduced temperature of the precursor softenings, $%
\Delta S_{\mathrm{power}}=$ $\left( E/E_{0}\right) ^{-1}-S_{bg}$, where $%
S_{bg}$ is obtained from the fits setting $A=0$. The exponents in the
legends are obtained from the linear fits of the log--log plots and coincide
with the $\kappa $ parameters of the corresponding fits. The deviations from
the power law at small reduced temperature depend largely on the first order
step of the modulus at the transition point between the cubic and tetragonal
phase and the very large softening regime. Particularly impressive is the
fact that BaTiO$_{3}$ closely follows a power law to $T_{\mathrm{C}}+360$~K,
the maximum temperature measured.

\section{Discussion}

The focus of the present analysis is the precursor softening in Ba$_{x}$Sr$%
_{1-x}$TiO$_{3}$\ beyond the classical critical regime, from few Kelvin
above $T_{\text{tr}}$\ up to the highest temperature reached in our
experiments. The major finding is that even in this huge temperature range
it is possible to describe the temperature dependence of the elastic moduli
in terms of a power law with exponent in line with predictions from models
of anharmonic phonon softening or, with somewhat less success, with a
Vogel-Fulcher expression. In both cases the fitting parameter evolve with
composition as expected from increasing order-disorder character from SrTiO$%
_{3}$\ to BaTiO$_{3}$.

The power-law and Vogel-Fulcher elastic softening theoretically converge for
small Vogel-Fulcher energies with power-law exponents near unity. We can
therefore define a rather strict "displacive limit" when the Vogel-Fulcher
energy vanishes, which is nearly the case for SrTiO$_{3}$. In all other
compounds we observe small but finite values of $E_{a}$ which indicate some
weak contributions of structural disorder. This component increases with
increasing Ba content, but always remains below the behaviour of a typical
order-disorder system. On the other hand, the power law fits show that the
anharmonic phonons coupled with the elastic moduli change their character,
in particular their dimensionality in the model in Ref. \cite{carpenter1998}%
. If a single branch flattens, the exponent $\kappa $ becomes $1.5$. If two
orthogonal branches flatten while the third remains relatively steep, we
expect $\kappa =$ $1$. Finally, if three orthogonal branches flatten, the
expected value is $\kappa =$ $0.5$. The experimental observations are
surprisingly close to these values for SrTiO$_{3}$\ while somewhat
exceed the lower limit for BaTiO$_{3}$. Simultaneously we observe that the
precursor temperature interval is smallest for SrTiO$_{3}$ while it remains
extremely large for Ba-rich materials with significant softening at least up
to 800~K. Similar exponents, but smaller precursor temperature intervals,
were previously observed in isostructural KMnF$_{3}$ and KMn$_{x}$Ca$_{1-x}$F%
$_{3}$ \cite{schranz2009, salje2009b, cao1988} with values of $\kappa $
ranging between $0.4$ and $1$. In PbSc$_{0.5}$Ta$_{0.5}$O$_{3}$ (PST) the
exponent is near 0.5 \cite{gan2019} with a smaller precursor interval, while
$\sim 1/3$\ has recently been found for structural fluctuations in cuprate
superconductors \cite{PSA22}.

An intriguing consequence of the precursor softening and the ensuing
potential heterogeneity in SrTiO$_{3}$ was discussed in \cite{bussmann2020}
where a strong coupling with the local carrier concentration was described.
It was found that even very small amounts of dopants can stabilize the soft
phonon branches. This is linked to the formation of polar nano-regions,
which grow in size with decreasing temperature. Such nano-regions would
modify the precursor exponent $\kappa $ and would also lead to an increase
of the VF activation energy. More generally \cite{bussmann2018} precursor
effects are a key to understand the fundamental aspects of the structural
phase transitions in perovskite structures which are unrelated to some
traditional concepts of the structural tolerance factors and small-cell DFT
calculations. The mixing of transition mechanisms is most obvious in BaTiO$%
_{3}$ and related materials. Their phase transitions are accompanied by
optic mode softening which remains incomplete near $T_{\text{tr}}$ due to
the first order nature of the phase transition \cite{harada1971}.
Simultaneously local probe measurements show an order-disorder behaviour
\cite{zalar2005, tai2004, ishidate1989}. The coexistence of both,
order/disorder and displacive mechanisms, has been widely discussed \cite%
{kraizman1995, stachiotti1993}. Migoni \textit{et al.} \cite{migoni1976} pointed out
that the most important ingredient of modelling of the soft mode activities
is the directional anisotropic core-shell coupling at the oxygen ion lattice
site, which is nonlinear with respect to the transition metal and harmonic
with respect to the A-site cation. This mirrors our observation that
dimensionality and hence the intrinsic anisotropy of the softening process
play a major role in the precursor effects of Ba$_{1-x}$Sr$_{x}$TiO$_{3}$.
We relate the difference between the end members for both model fits to the
different phonon dispersions as calculated by Busmann-Holder and co-workers
\cite{bussmann2018}, who found that the transverse acoustic zone boundary TA
mode couples strongly to the optic mode at finite momentum. The TA mode is
presumed to be the origin of finite size precursors and polar nanoregions.
This mode is less temperature dependent than the soft TO mode, which softens
more strongly in BaTiO$_{3}$ and less in SrTiO$_{3}$. A crossing of the zone
boundary and the zone centre mode with decreasing temperature occurs in BaTiO%
$_{3}$ but not in SrTiO$_{3}$. The avoided mode crossing induces precursor
dynamics at small momenta which are less developed in BaTiO$_{3}$ where the
optic mode softens over the full momentum space. These authors already
argued that the softening of the optic mode in BaTiO$_{3}$ for all momenta
is reminiscent of the breather type excitation and a signature for the
formation of polar nanoregions far above the actual phase transition, which
mirrors our findings.

There appears no definite onset for the softening, because all compositions
can be well fitted up to $4T_{\mathrm{C}}$\ or the highest temperature
reached experimentally with a power law or VF expression, without a
characteristic temperature. This was expected for the displacive transition
of non-polar SrTiO$_{3}$, but not for Ba-rich compositions and especially
BaTiO$_{3}$, where characteristic temperatures have been identified below
which polar nanoclusters start forming, i.e. the Burns temperature $%
T_{d}\sim $\ 550--590~K and $T^{\ast }\sim $\ 500~K \cite%
{BD82,DPK10,MJB04,KT10,PKS12,SCN13}. The experimental signatures of such
onsets or crossover temperatures are deviations from linearity in the
temperature dependence of the refraction index \cite{BD82}, of thermal
expansion \cite{MJB04}, of the elastic constants at very high frequency \cite%
{KT10}, or maximum in the elastic moduli at MHz frequencies \cite{SCN13},
bursts of Acoustic Emission \cite{DPK10}, and a subtle change in the
exponent of the power law describing the Second Harmonic Generation \cite%
{PKS12}. Below $T_{d}$\ nanoregions with static average displacements have
been observed to develop with STEM and Raman spectroscopy \cite{bencan2021}.

\begin{figure}[t]
\begin{center}
\includegraphics[width=8.5cm]{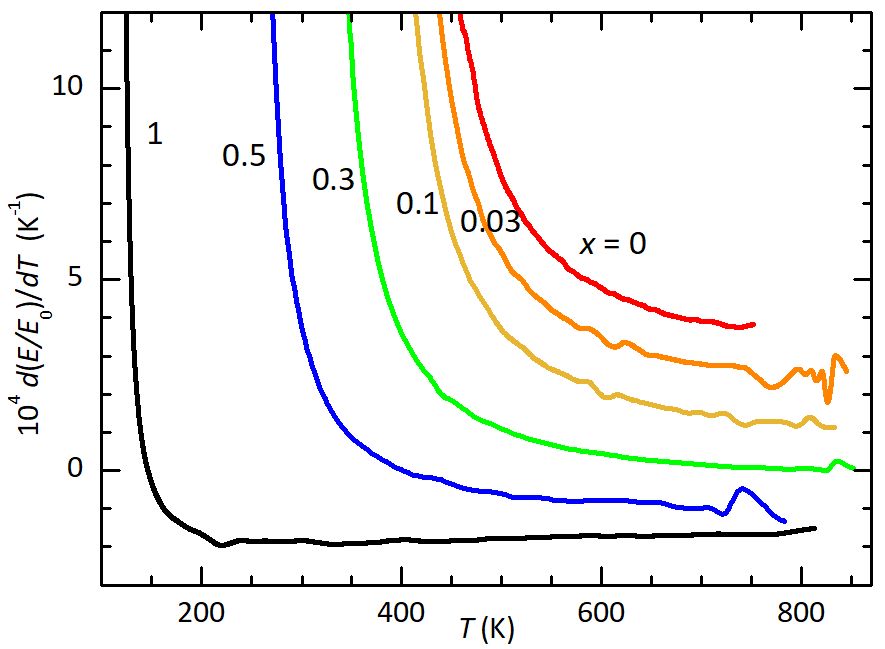}
\end{center}
\caption{Temperature derivatives of the Young's moduli of Ba$_{x}$Sr$_{1-x}$%
TiO$_{3}$. For clarity the curves are shifted of $10^{-4}$~K$^{-1}$ with
respect to each other. The larger noise at the highest temperature in some
curves is due to the fact that the measurements were very fast there (up to
7~K/min) in order to limit possible losses of O in high vacuum.}
\label{fig-dMdT}
\end{figure}

Our data and their analyses do not show any evidence of such characteristic
temperatures up to the highest we reached. This is better seen in the
temperature derivatives of the moduli, shown in Fig. \ref{fig-dMdT}, where
there is no evidence of inflections of the $E\left( T\right) $\ curves.

It appears possible that in Ba$_{x}$Sr$_{1-x}$TiO$_{3}$\ (away from the
relaxor compositions $x\sim 0.15$) the measured $T_{d}$\ may define the
onset temperature range below which some precursor phenomena were detected,
while such temperatures seem not to appear in elastic softening. There are
many theoretical models to account for elastic precursor softening,
including the appearance of polar nano-clusters, local intrinsic
correlations of precursor displacements \cite{ZMB22} etc. while,
surprisingly, the quantitative temperature evolution of the excess softening
follows simple power-law (or Vogel-Fulcher) dynamics. This observation also
reveals that the bare elastic moduli (when the precursor softening is
eliminated) are much bigger than measured even at very high temperatures. As
an example, with power-law fitting, at the highest temperature 750~K reached
in BaTiO$_{3}$ the bare elastic modulus is 34\% larger than the
experimentally observed value, and at 840 K the bare modulus of Ba$_{0.03}$Sr%
$_{097}$TiO$_{3}$ is 22\% larger than the experimental value, so that BaTiO$%
_{3}$ would be even stiffer than SrTiO$_{3}$ in the absence of precursor
softening.

\section{Conclusions}

Elastic precursor effects are common in ferroelastic materials. An
increasing number of publications report such effects although with little
or no numerical analysis of the observed temperature dependencies. We
proposed two frameworks to analyse such effects by power law dependence and
a Vogel-Fulcher dynamics. It is very likely that these approaches are indeed
universal. We urge the community to analyse such effects in the paraelastic
phase and/or paraelectric phase because several applications may be based
not only on domain boundaries \cite{nataf2020} but also on the dynamic
effects in the high symmetry phase. To demonstrate this behaviour, we
reported in this paper the precursor softening of two prototypical materials
and their mixed crystals. We found a smooth crossover between a displacive
and a (partially) order-disorder system so that the precursor dynamics can
be tailored according to specific applications. We presume that other
structural imperfections, such as oxygen defects, will contribute similarly
to local structural disorder and enhance the BaTiO$_{3}$-type elastic
softening with extremely wide temperature intervals and low exponents $%
\kappa $.

\section*{Acknowledgments}

The authors thank Vincenzo Buscaglia (CNR-ICMATE)\ for supplying the ceramic
samples of SrTiO$_{3}$ and Ba$_{0.5}$Sr$_{0.5}$TiO$_{3}$ and Massimiliano Paolo
Latino (CNR-ISM) for his technical assistance. EKHS is grateful
to EPSRC for financial support (EP/P024904/1). PSSJ and MV are grateful to the
Brazilian funding agencies for financial support: S\~{a}o Paulo State Research
Foundation FAPESP (grants \#2012/08457-7 and \#2022/08030-5) and National
Council for Scientific and Technological Development (CNPq) grant \#304144/2021-5.

\bibliographystyle{unsrt}
\bibliography{refs}

\end{document}